# A VOLCANIC HYDROGEN HABITABLE ZONE

Ramses M. Ramirez[1,2], and Lisa Kaltenegger[1,3]

[1]Carl Sagan Institute, Cornell University, Ithaca, NY
[2]Cornell Center of Astrophysics and Planetary Science, Cornell University, Ithaca, NY
[3]Department of Astronomy, Cornell University, Ithaca, NY

Corresponding author - Ramses Ramirez email: rmr277@cornell.edu, phone 480-296-6477

## ABSTRACT

The classical habitable zone is the circular region around a star in which liquid water could exist on the surface of a rocky planet. The outer edge of the traditional $N_2$-$CO_2$-$H_2O$ habitable zone (HZ) extends out to nearly ~1.7 AU in our Solar System, beyond which condensation and scattering by $CO_2$ outstrips its greenhouse capacity. Here, we show that volcanic outgassing of atmospheric $H_2$ on a planet near the outer edge can extend the habitable zone out to ~2.4 AU in our solar system. This wider volcanic hydrogen habitable zone ($N_2$-$CO_2$-$H_2O$-$H_2$) can be sustained as long as volcanic $H_2$ output offsets its escape from the top of the atmosphere. We use a single-column radiative-convective climate model to compute the HZ limits of this volcanic hydrogen habitable zone for hydrogen concentrations between 1% and 50%, assuming diffusion-limited atmospheric escape. At a hydrogen concentration of 50%, the effective stellar flux required to support the outer edge decreases by ~35% to 60% for M - A stars. The corresponding orbital distances increase by ~30% to 60%. The inner edge of this HZ only moves out ~0.1 to 4% relative to the classical HZ because $H_2$ warming is reduced in dense $H_2O$ atmospheres. The atmospheric scale heights of such volcanic $H_2$ atmospheres near the outer edge of the HZ also increase, facilitating remote detection of atmospheric signatures.

## 1. INTRODUCTION

The habitable zone (HZ) is the circular region around a star(s) in which liquid water could be stable on a rocky planet's surface (e.g. Kasting et al., 1993, Kaltenegger & Haghighipour 2013, Haghighipour & Kaltenegger 2013) and possible atmospheric biosignatures may be detected (e.g. Kaltenegger et al. 2010). The classical $N_2$-$CO_2$-$H_2O$ HZ is defined by the greenhouse effect of $CO_2$ and $H_2O$ vapor. The outer edge, the $CO_2$ maximum greenhouse limit, is defined as the distance beyond which condensation and scattering by $CO_2$ outstrips its greenhouse capacity. The inner edge is defined where mean surface temperatures exceed the critical point of water (~647 K, 220 bar), triggering a runaway greenhouse that leads to rapid water loss to space on very short timescales (see Kasting et al. 1993 for details). Additional greenhouse gases could further extend the HZ. It had been suggested that young orphan planets unbound to their host stars could accrete massive primordial hydrogen envelopes (Stevenson 1999). Models show that a super-Earth with a 40 bar primordial hydrogen atmosphere could maintain above-freezing mean surface temperatures out to 10 AU from a solar-type G-star (Pierrehumbert and Gaidos 2011). The potency of this greenhouse effect arises from collision-induced absorption (CIA) caused by self-broadening from $H_2$-$H_2$ collisions, allowing $H_2$ to function noncondensibly out to great distances (Pierrehumbert and Gaidos, 2011). For comparison, condensation and scattering effects at high $CO_2$ partial pressures limit the outer edge of the classical HZ in our solar system to ~1.7 AU (e.g. Kasting et al., 1993).

However, maintaining large amounts of primordial hydrogen over geological timescales is difficult (Pierrehumbert and Gaidos, 2011). Hydrogen is a very light greenhouse gas, and without a continuous source, hydrodynamic escape to space would strip a 5 Earth-mass habitable zone rocky planet with 50 bars of primordial atmospheric $H_2$ within just a few million years. Primordial hydrogen lifetimes in atmospheric models only extend to ~100 million years at $H_2$ partial pressures of several hundred bar (Wordsworth, 2012).

Another way to generate atmospheric hydrogen on terrestrial planets is through volcanism. Paleoclimate studies of early Earth and Mars show that volcanism could have outpaced escape of $H_2$ on both planets, maintaining clement conditions with modest $H_2$ partial pressures below 1 bar (Wordsworth and Pierrehumbert, 2013; Ramirez et al., 2014). Reduced mantle conditions could have favored enhanced outgassing of $H_2$ over longer timescales. In these models hydrogen is not the major atmospheric constituent and is continuously replenished by volcanism that offsets losses to space. In these volcanic hydrogen atmospheres, it is *foreign-broadening* by the remaining background atmosphere that excites roto-translational bands within the hydrogen, increasing greenhouse warming in spectral regions where $CO_2$ and $H_2O$ absorb poorly (Ramirez et al. 2014).

In this work, we assess the impact that foreign broadening of volcanic hydrogen in a planet's atmosphere has on the classical $N_2$-$CO_2$-$H_2O$ habitable zone (Kasting et al., 1993; Kopparapu et al., 2013). We derive the limits of this volcanic-hydrogen HZ ($N_2$-$CO_2$-$H_2O$-$H_2$) for host stars with $T_{eff}$ ranging from 2,600K to 10,000 K.

## 2. METHODOLOGY

*2.1 Climate modeling procedures*

As in previous studies of the HZ (e.g. Kopparapu et al., 2013; Ramirez and Kaltenegger, 2014; Ramirez and Kaltenegger, 2016) we used a 1-D radiative-convective climate model to compute habitable zone boundaries for stars of stellar effective temperatures ($T_{eff}$) ranging from 2,600K to 10,000K (Ramirez and Kaltenegger, 2016). The model uses correlated-k coefficients to parameterize absorption by $H_2O$ and $CO_2$ across 38 solar intervals and 55 infrared intervals (e.g. Kopparapu et al. 2013 for details). We model $CO_2$-$H_2$ collision-induced absorption (CIA) using recently computed ab initio calculations (Wordsworth et al., 2017), $CO_2$-$CO_2$ CIA using established parameterizations (Gruszka and Borysow, 1997; Gruszka and Borysow, 1998; Baranov et al., 2004), self-broadening of $H_2$-$H_2$ pairs (Borysow, 2002) and $N_2$ foreign-broadening (see Ramirez et al. 2014 for details).

We use inverse calculations, where the surface temperature is specified and the solar flux required to maintain it is calculated, to determine the limits of the HZ (following e.g., Kasting et al., 1993; Kopparapu et al., 2013). For the outer edge HZ calculations, the mean surface temperature is fixed at 273 K, above which liquid water is stable on a planetary surface. Stratospheric temperatures are fixed at 154 K, which is representative for a planet near the outer edge of the habitable zone (e.g. Kopparapu et al. 2013). For the inner edge, the stratospheric temperature is 200 K and the surface temperature is gradually increased from a starting temperature of 200 K, which simulates pushing the planet closer and closer to the star until a runaway greenhouse is triggered (e.g. Kasting et al., 1993).

The input of hydrogen from volcanic sources in our models is balanced by its escape to space. Actual hydrogen escape rates will be planet-specific, therefore we assume that hydrogen escapes at the diffusion limit, which is the fastest that H can escape at the concentrations considered (e.g Walker, 1977). Thus, our results may overestimate hydrogen escape and thus provide a lower bound on H concentrations. As with previous HZ studies, we assume planets support plate tectonics, including a carbonate-silicate cycle capable of maintaining habitability over long timescales (e.g. Kasting et al., 1993, Ramirez & Kaltenegger 2016). We also assume highly-reduced mantle conditions on terrestrial planets, as suggested from meteoritic evidence from early Mars (e.g. Grott et al., 2011). Hydrogen sources and sinks need to balance to support a continuous hydrogen concentration. Our model planet has the same outgassing rates per unit area as those on our present day planet based on the inference that the heat flux of early Mars could have been similar to that of modern Earth (Montési and Zuber, 2003; Ramirez et al., 2014). This allows us to constrain typical $H_2$ concentrations for reducing mantles of outer edge planets to ~ 1 – 30% (Ramirez et al., 2014; Batalha et al., 2015). For comparison we include one extreme concentration of 50% (assuming even higher outgassing rates or less efficient hydrogen escape) intended to simulate potentially higher outgassing rates (e.g. on hypothetical planets with unusually high geothermal fluxes).

Each of our model atmospheres contains 1 bar of $N_2$ and $H_2$ concentrations of 1, 5, 10, 20, 30, and 50%, respectively, of the dry ($N_2$+$CO_2$+$H_2$) atmosphere. Oxygen has only a weak greenhouse effect and is replaced by $N_2$ in these calculations (following Kasting et al. 1993). For the outer edge calculations we vary the $CO_2$ partial pressure from $1\times10^{-2}$ to 34.7 bar, the saturation $CO_2$ partial pressure at 273K. $CH_4$ is not expected to build up to appreciable concentrations in $CO_2$-$H_2$ atmospheres because of its short photochemical lifetime when $H_2$ is not the major constituent (Bains et al. 2014). Recent work shows that $CH_4$ would likely oxidize to $CO_2$ in warm $CO_2$-$H_2$ atmospheres (Batalha et al., 2015).

At the inner edge, our volcanic-$H_2$ atmospheres contain 1 bar of $N_2$ and 330 ppm $CO_2$ (following Kasting et al., 1993). As in the case of the outer edge, we model $H_2$ concentrations that are 1, 5, 10, 20, 30 and 50% of the dry ($N_2$+$CO_2$+$H_2$) atmosphere. We gradually increase planetary surface temperatures from 200K to the critical point of water (647.1K, 220.6 bars $H_2O$), which leads to the runaway greenhouse and rapid evaporation of the surface water inventory (e.g., Kasting et al., 1993). The high atmospheric $H_2O$ mole fraction at the inner edge reduces the warming effect of $H_2$. Note that the $H_2$ amount is scaled to the dry atmosphere because volcanism is not expected to increase as $H_2O$ vapor mixing ratios increase.

In keeping with previous work (see e.g. Kopparapu et al. 2013) all model atmospheres are fully-saturated: A moist $CO_2$ adiabat is followed in the upper atmosphere for the outer edge calculations. A moist $H_2O$ adiabat is followed in the upper atmosphere for the inner edge model calculations. We also employ six solar zenith angles and a surface albedo of 0.31 to mimic the combined clouds and surface reflection, which reproduces the mean surface temperature for present Earth (ibid). Because there is no self-consistent cloud model for 1D HZ studies (see Zsoms et al. 2012), surface and cloud albedo are held constant for all simulations (following previous 1-D HZ studies, see e.g. Kasting et al., 1993; Kopparapu et al., 2013), even though cloud feedback is expected to change for atmospheric compositions different from those of present-day Earth. However, our understanding of clouds is still evolving and their applicability to water-rich atmospheres (i.e. inner edge of the HZ) is not fully understood yet. Note that overall planetary albedo changes as the atmospheric composition is varied.

## 3. RESULTS

Adding volcanic hydrogen moves both the inner and the outer edges of the classical HZ outward and widens it. The incident stellar fluxes ($S_{eff}$) corresponding to the outer edges of the classical and new volcanic-$H_2$ HZ, are shown in fig.1. For comparison, an alternative HZ limit that is not based on atmospheric models (i.e. classical HZ), but on empirical observations of our solar system, is also shown in figs. 1-2. The inner edge of this empirical HZ is defined by the stellar flux received by Venus when we can exclude the possibility that it had standing water on the surface (about 1Gyr ago), equivalent to a stellar flux of $S_{eff}$=1.77 (Kasting et al.1993). The outer edge is defined by the stellar flux that Mars received at the time that it may have had stable water on its surface (about 3.8 Gyr ago), which corresponds to $S_{eff}$=0.32 (ibid). The incident flux in both figures is normalized to that received by Earth's orbit (~1360 W/m$^2$), $S_0$.

The incident stellar flux that is needed to maintain liquid water surface temperatures at the outer edge of our solar system decreases from 0.3576 to 0.267 (25% decrease) and to 0.201 (44% decrease) with 5% and 30% $H_2$, respectively (fig.1). This corresponds to increases in the orbital distance of the maximum greenhouse limit of $CO_2$ at the outer edge HZ from 1.67 AU to 1.94 AU and 2.23 AU, respectively (16% and 34% increases, respectively; fig.2). With 50% $H_2$ the limits extend farther, corresponding to an $S_{eff}$ of 0.172 (52% increase) and an outer edge distance of 2.4 AU (44% increase) in our own Solar System. Overall, the $S_{eff}$ required to support the outer edge decreases by a maximum of ~35 – 60% (at 50% $H_2$) for M - A star types, resulting in respective distance increases of ~30 to 60% (fig. 2).

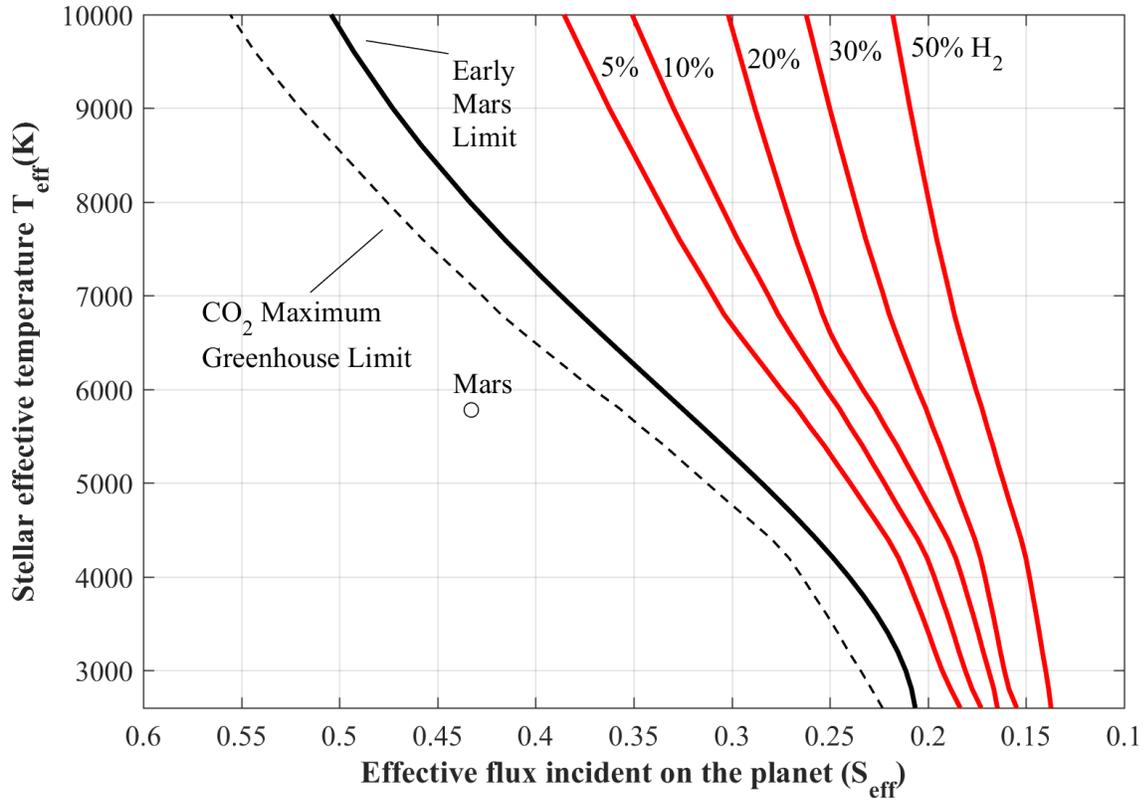

**Figure 1:** Effective stellar temperature versus incident stellar flux ($S_{eff}$) for the outer edge. The $CO_2$ maximum greenhouse limit (dashed) is shown along with the empirical outer edge (solid black) and outer edge limits containing 5%, 10%, 20%, 30%, and 50% $H_2$ (red solid)

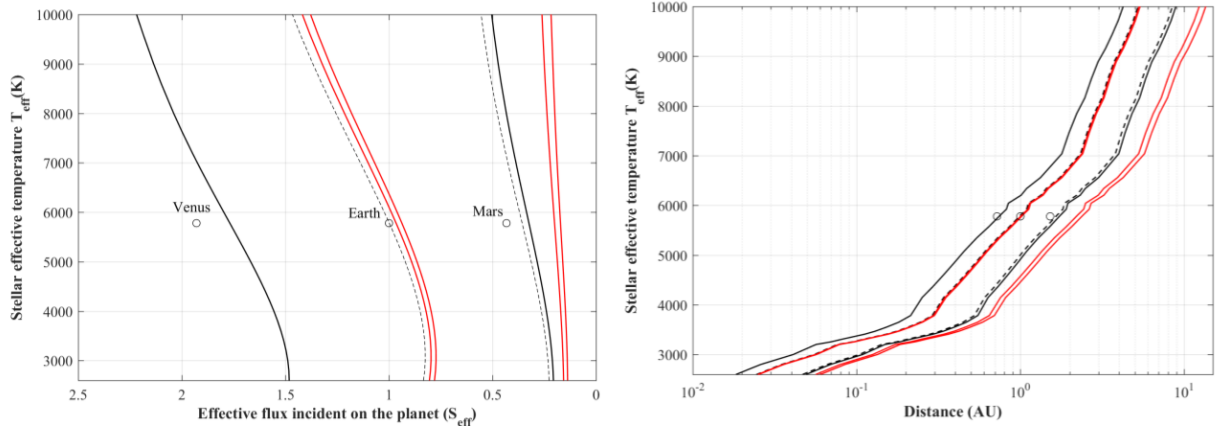

**Figure 2:** Effective stellar temperature versus (left) incident stellar flux ($S_{eff}$) and (right) orbital distance for the classical (dashed), empirical (solid black), and volcanic hydrogen (solid red) habitable zone. The red curves contain 30% and 50% $H_2$, respectively. As shown in Kopparapu et al. (2013), Earth appears near the classical inner edge because of the generic model assumption of 100% relative humidity. With subsaturation, Earth is well within the classical HZ (e.g., Leconte et al. 2013)

The volcanic hydrogen HZ boundaries can be calculated using equation 1 with constants from Table 1 and Table 2:

$$S_{eff} = S_{eff(sun)} + aT_* + bT_*^2 + cT_*^3 + dT_*^4 \quad (1)$$

where $T_*$ is equal to $T_{eff} - 5780K$ and $S_{eff(sun)}$ are the fluxes computed for our solar system (see Tables 1-2). The corresponding orbital distances of the HZ can be calculated using equation 2 (fig.2):

$$d = \sqrt{\frac{L/L_{sun}}{S_{eff}}} \quad (2)$$

where $L/L_{sun}$ is the stellar luminosity in solar units and $d$ is the orbital distance in AU. Planetary mass that is linked to planetary model surface pressure via gravity and planetary radius have only a small effect on the HZ limits (Kopparapu et al. 2014).

Although the inner edge of the HZ is also affected by the addition of $H_2$, the runaway greenhouse limit distance (e.g. Kasting et al., 1993) only moves outward by ~0.1% to 4% for 1% $H_2$ and 50% $H_2$ (fig. 2), respectively, because of the dense water vapor atmosphere at the inner edge. At temperatures above ~300 K water vapor absorption begins to mask hydrogen collision-induced absorption, decreasing its effectiveness.

The greenhouse effect of $CO_2$ increases with higher $pCO_2$ to a maximum value (fig. 3), corresponding to the maximum greenhouse effect. However, as the hydrogen concentration increases, the $CO_2$ partial pressures associated with the outer edge occur at lower $S_{eff}$ and larger orbital distances. This occurs because cooling from both $CO_2$ condensation and scattering eventually outpace the combined greenhouse effect of $CO_2$ and $H_2$ farther from the host star.

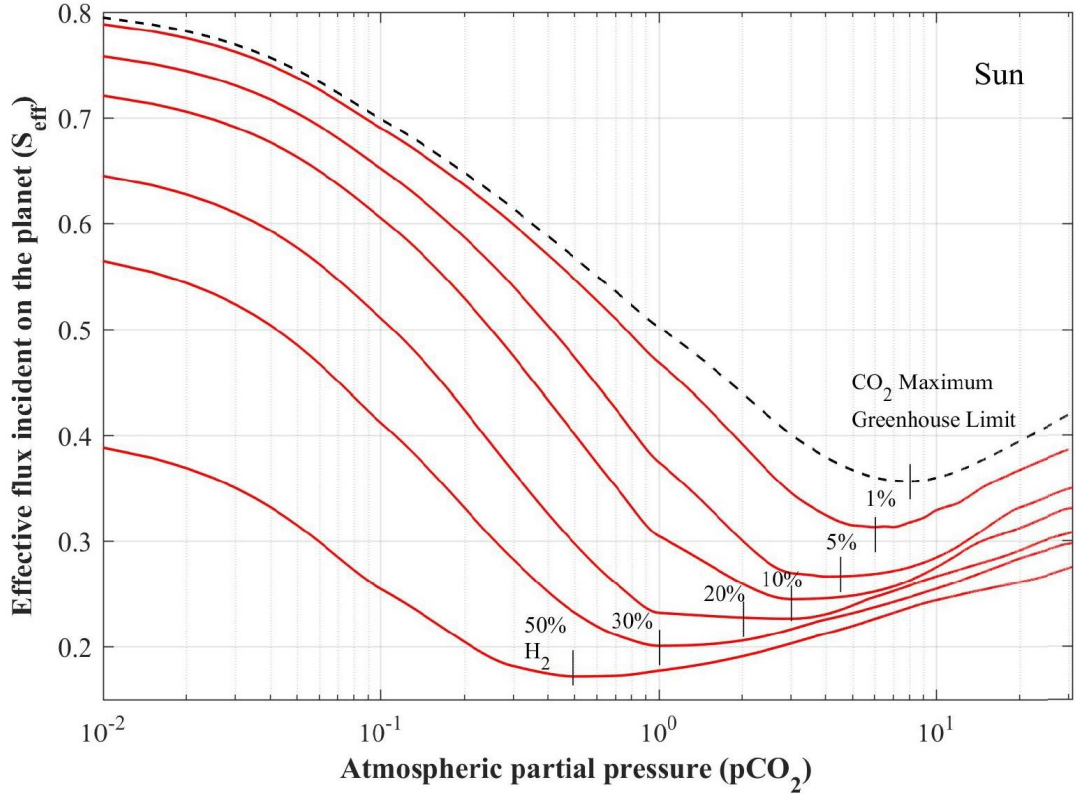

**Figure 3:** Dependence of the incident stellar flux ($S_{eff}$) on atmospheric $CO_2$ partial pressure ($pCO_2$) for the solar system outer edge. The $pCO_2$ values corresponding to the maximum greenhouse for atmospheres with hydrogen (red curves) and without (dashed curve) are shown as vertical lines for 0% to 50% $H_2$ (see text).

About 8 bars of $CO_2$ are required to sustain surface temperatures above freezing at the $CO_2$ maximum greenhouse limit of the classical HZ in our solar system (Kasting et al., 1993; Kopparapu et al., 2013). When hydrogen is added to the planetary atmosphere, the $CO_2$ partial pressure required at the outer edge of the volcanic-$H_2$ HZ decreases to 6 bar with 1% $H_2$, and to 1 bar at 30% $H_2$ (fig. 3). The corresponding $H_2$ partial pressures (with 1 bar $N_2$) are 0.071 bar and 0.857 bar respectively. At 50% $H_2$, the $CO_2$ and $H_2$ partial pressures at the outer edge of the volcanic-hydrogen HZ are 0.5 and 1.5 bars, respectively.

The stellar energy distribution of host stars also determines what $CO_2$ concentration the maximum $CO_2$ greenhouse effect is reached. For cooler stars, increased near-infrared absorption and reduced scattering lower this $CO_2$ concentration (e.g. Kasting et al., 1993). For a cool M6 star ($T_{eff}$ ~ 3000 K), the $CO_2$ partial pressure at the outer edge of the classical HZ is 15 bar. Adding $H_2$ reduces the required amount of $CO_2$ at the outer edge to 9 bar and 1 bar for 1% and 30% $H_2$, respectively. The corresponding $H_2$ partial pressures are 0.1 bar and 0.857 bar.

Atmospheric scale height increases with the addition of hydrogen. For example, adding 30% $H_2$ to the atmosphere of a habitable planet located in our solar system's outer edge (1 bar $N_2$ and 8 bar $CO_2$) increases its atmospheric scale height by over a factor of 1.5 due to the decreased atmospheric mean molecular mass, assuming a similar temperature structure. In addition, such $H_2$-rich planets at the outer edge of the HZ also have less $CO_2$ in their atmospheres, reducing the optical density as compared to those $CO_2$ atmospheres that do not contain hydrogen. These two effects could facilitate the detection of atmospheric spectral features with the next generation of telescopes, making such planets very interesting targets for JWST and the E-ELTs in the search for life.

## 4. DISCUSSION

Although the present day mantle of Earth is relatively oxidized, it is generally assumed that the mantles in terrestrial planets, including Earth, start reduced (i.e. oxygen-poor), which may lead to conditions that promote high levels of $H_2$ outgassing. One of the main arguments for a reduced early mantle is that Earth (and possibly other terrestrial planets) must have entered a magma ocean stage during accretion, which would have acted as a surficial oxygen sink (e.g., Scaillet and Gaillard, 2011). Afterwards, planetary mantles are thought to become oxidized over time, either gradually or in a stepwise fashion (Wood et al., 2006). Mantle oxidation on Earth may have occurred rather quickly, perhaps only ~100 million years after core-mantle differentiation (Trail et al., 2011).

Mantle oxidation may have taken longer on Mars though. High $H_2$ outgassing on early Mars is inferred from Martian meteorites suggesting a highly reduced ancient Martian mantle (Stanley et al., 2011; Grott et al., 2011). If early Mars was kept warm by atmospheric $H_2$ up until ~3.6 Ga, this would suggest reducing mantle conditions had persisted for ~1 billion years. Supporting this view, geochemical results from the ~4 billion year old meteorites, ALH84001 and NWA Black Beauty, infer a reduced mantle for at least ~0.5 billion years (e.g. Grott et al., 2011).

These solar system examples warrant some discussion of how mantle oxidation may generally proceed on planets. One idea suggests lower mantle pressures of larger terrestrial bodies (e.g. Earth) are sufficiently high to convert iron(II) oxide to iron metal and iron(III) oxide, oxidizing the mantle during accretion (Wade and Wood, 2005). According to this notion, smaller planets like Mars would not have generated internal pressures sufficiently high for mantle oxidation, in agreement with the meteorite evidence. This may suggest that atmospheric $H_2$ may not easily build up on more massive planets. However, this may happen in a number of ways. First, these calculations assume that $H_2$ escapes at the diffusion limit, but $H_2$ concentrations can potentially build up if escape rates (including hydrodynamic and Jean's) are lower than this limit, as had been suggested for the early Earth (Tian et al. 2005). Hydrogen retention will also be favored on planets with stronger magnetic fields or on those with higher outgassing rates per unit area.

Moreover, it is possible that $H_2$-rich atmospheres may be longer-lived on super-Earths. First, their enhanced gravity results in reduced $H_2$ escape rates, promoting hydrogen retention (e.g. Pierrehumbert and Gaidos, 2011). Secondly, mantle oxidation may be slowed as a result, increasing the lifetime of reduced mantle conditions. Larger planets may have higher outgassing rates, possibly leading to a feedback in which higher outgassing rates enhance weathering rates and vice versa (e.g. Heller 2015). Plus, interior heat is retained longer on super-Earths, ensuring that plate tectonics and volcanism both last (ibid).

Hydrogen retention is also favored for outer edge planets. Hamano et al. (2013) suggest that selective hydrodynamic escape of H over O during accretion could lead to early mantle oxidation. However, outer edge planets would be less susceptible to this effect because corresponding EUV fluxes are much lower than in the inner solar system. Given that hydrodynamic escape rates are $\propto 1/(d(AU))^2$, and assuming energy-limited escape (e.g. Ramirez and Kaltenegger, 2014), H escape rates (all else equal) on a young accreting Earth-mass planet orbiting the early Sun at 1.7 AU would have been ~ a factor of ~3 – 6 lower than on early Earth or Venus. These loss rates would decrease further by $1/30 - 1/60$ on a 10-Earth mass super-Earth. It is also possible that accreting planets incorporate more hydrogen because the protoplanetary disk itself is enriched in the element. In addition, because stellar fluxes are lower for these distant planets, magma ocean durations should decrease as well, further weakening escape (Hamano et al. 2013). All of these factors could more than offset the factor of ~15 – 40 relatively higher $H_2$ outgassing rates inferred for a highly reduced mantle (Ramirez et al., 2014). Thus, sizeable hydrogen inventories on larger terrestrial planets located near the outer edge may be sustainable as long as mantle conditions support conditions that favor hydrogen retention over escape to space.

Certain atmospheric spectral features, including $N_2O$ and $NH_3$, which can, but do not have to be produced biotically, could be detected in $H_2$-dominated atmospheres (Seager et al., 2013; Baines et al., 2014). Such volcanic-hydrogen atmospheres may also be able to evolve methane-based photosynthesis (Bains et al., 2014). Distinguishing biosignatures from abiotic sources in such atmospheres will be challenging. $NH_3$ can be formed abiotically through reaction of $N_2$ and $H_2$ in hydrothermal vents on planets with reducing mantles (Kasting et al., 2014). $N_2O$ can be formed a number of ways including through atmospheric shock from meteoritic fall-in, lightning, UV radiation (e.g. Ramirez, 2016) and through solar flare interactions with the magnetosphere (Airapetian et al., 2016). Thus, future biosignature studies should focus on modeling biotic and abiotic sources for these gases in thin volcanic-hydrogen atmospheres.

Lastly, earlier studies suggesting that backscattering from $CO_2$ clouds could move the outer edge beyond 2 AU (Mischna et al., 2000) is challenged by new work indicating that the greenhouse effect of $CO_2$ clouds may be greatly overstated (Kitzmann et al., 2016; 2017). Even should warming from $CO_2$ clouds be relatively ineffective, volcanically outgassed $H_2$ in some stellar systems may compensate and keep the HZ similarly wide for geologically significant timescales.

## 5. CONCLUSION

We have calculated the limits of the volcanic-$H_2$ HZ for $H_2$ concentrations ranging from 1% to 50%. We find that the addition of 30% $H_2$ can extend the outer edge in our solar system to ~2.23 AU (2.4 AU with 50% $H_2$). The orbital distance of the outer edge of the volcanic-hydrogen HZ moves outward by as much as ~60% as compared to the classical HZ for A to M type main sequence Stars. The inner edge only moves out by ~0.1% to 4%.

Reduced mantle conditions that promote hydrogen outgassing on small terrestrial planets may be maintained for long timescales (>0.5 Gyr). Atmospheric $H_2$ may be retained on larger terrestrial planets, including super-Earths, through higher gravity and potentially stronger magnetic fields. H retention is also generally favored in planets with higher outgassing rates, and if H$_2$ escapes more slowly than the diffusion limit. Hydrodynamic escape rates are also lower for planets farther away from the host star, mitigating the selective escape of H over O. Generally, hydrogen-rich atmospheres composed of N$_2$, CO$_2$ and H$_2$O can be maintained as long as volcanic outgassing of H$_2$ outpaces its escape.

Moreover, the addition of H$_2$ increases the atmospheric scale height making super-Earths near the volcanic-H$_2$ HZ outer edge interesting observational targets. Near-future missions like JWST and the ELTs could potentially detect atmospheric features, including biosignatures, on such planets.

**Table 1.** Coefficients to calculate the outer edge of the volcanic hydrogen HZ

| Constant | 1% H$_2$ | 5% H$_2$ | 10% H$_2$ | 20% H$_2$ | 30% H$_2$ | 50% H$_2$ |
|---|---|---|---|---|---|---|
| $S_{eff(Sun)}$ | 0.3129 | 0.2671 | 0.2452 | 0.2271 | 0.2009 | 0.1723 |
| A | 4.6597x10$^{-05}$ | 3.4301x10$^{-5}$ | 2.9373x10$^{-5}$ | 2.5976x10$^{-5}$ | 1.7821x10$^{-5}$ | 1.4029x10$^{-05}$ |
| B | 3.7621x10$^{-10}$ | 1.2999x10$^{-10}$ | 2.3132x10$^{-10}$ | -1.0703x10$^{-9}$ | -6.9474x10$^{-11}$ | -3.2318x10$^{-10}$ |
| C | -9.332x10$^{-13}$ | -6.6514x10$^{-13}$ | -5.1625x10$^{-13}$ | -6.8630x10$^{-13}$ | -3.1860x10$^{-13}$ | -2.8689x10$^{-13}$ |
| D | 9.4571x10$^{-17}$ | 6.3759x10$^{-17}$ | 4.9119x10$^{-17}$ | 1.1099x10$^{-16}$ | 3.3181x10$^{-17}$ | 4.2926x10$^{-17}$ |

**Table 2.** Coefficients to calculate the inner edge of the volcanic hydrogen HZ

| Constant | 1% H$_2$ | 5% H$_2$ | 10% H$_2$ | 20% H$_2$ | 30% H$_2$ | 50% H$_2$ |
|---|---|---|---|---|---|---|
| $S_{eff(Sun)}$ | 0.9975 | 0.9946 | 0.9884 | 0.9775 | 0.9655 | 0.9377 |
| a | 1.0708x10$^{-4}$ | 1.0644x10$^{-4}$ | 1.0613x10$^{-4}$ | 1.0501x10$^{-4}$ | 1.0383x10$^{-4}$ | 1.0123x10$^{-4}$ |
| b | 8.5594x10$^{-9}$ | 7.4458x10$^{-9}$ | 8.4633x10$^{-9}$ | 8.3633x10$^{-9}$ | 8.2456x10$^{-9}$ | 7.9833x10$^{-9}$ |
| c | -2.3600x10$^{-12}$ | -2.4365x10$^{-12}$ | -2.3376x10$^{-12}$ | -2.3106x10$^{-12}$ | -2.2837x10$^{-12}$ | -2.2204x10$^{-12}$ |
| d | 1.2683x10$^{-16}$ | 2.0284x10$^{-16}$ | 1.2619x10$^{-16}$ | 1.2447x10$^{-16}$ | 1.2364x10$^{-16}$ | 1.2098x10$^{-16}$ |

ACKNOWLEDGEMENTS

The authors would like to thank Robin Wordsworth for kindly providing his $CO_2$-$H_2$ collision-induced absorption data. We also thank James F. Kasting for the helpful discussions. The authors acknowledge support by the Simons Foundation (SCOL # 290357, Kaltenegger) and the Carl Sagan Institute.

**REFERENCES**

Airapetian, V. S., Glocer, A., Gronoff, G., Hébrard, E., & Danchi, W. 2016, *NatGe.*, 9, 452- 455

Bains, W., Seager, S., & Zsom, A. (2014). *Life*, *4*(4), 716-744.

Baranov, Y. I., Lafferty, W. J., & Fraser, G. T. 2004, *Journal of molecular spectroscopy*, *228*,2, 432-440.

Batalha, N., Domagal-Goldman, S. D., Ramirez, R., & Kasting, J. F. 2015, *Icarus*, *258*, 337-349.

Borysow, A. 2002, *A&A*, 390, 2, 779-782.

Grott, M., Morschhauser, A., Breuer, D., & Hauber, E. 2011, *Earth and Planetary Science Letters*, *308*, 3, 391-400.

Gruszka, M., & Borysow, A. 1997, *Icarus*, *129*, 1, 172-177.

Gruszka, M. Borysow, A. *Molecular physics* 93.6 (1998): 1007-1016.

Haghighipour N. & Kaltenegger, L, Habitability of Binary Systems II: P-type binaries, *ApJ*, 777, 2, 166, 13 pp., 2013

Hamano, K., Abe, Y., & Genda, H. 2013, *Natur*, *497*, 7451, 607-610.

Heller, René, 2015. *Scientific American* 312,1, 32-39.

Kaltenegger, L. & Haghighipour N., Habitability of Binary Systems I: S-type binaries, *ApJ*, 777, 2, 165, 11 pp., 2013

Kaltenegger, L., Selsis, F., Fridlund, M. et al. (2010). *AsBio*, *10*(1), 89-102

Kasting, J. F., Kopparapu, R., Ramirez, R. M., & Harman, C. E. 2014, *PNAS*, *111*, 35, 12641-12646.

Kasting, J.F., Whitmire, D.P., Reynolds, R.T., 1993, *Icarus* 101 108-128

Kitzmann, D. 2016, *ApJL*, *817*, 2, L18.

Kitzmann, D. (2017). *arXiv preprint arXiv:1701.07513*.

Kopparapu, R. K., Ramirez, R., Kasting, J. F., Eymet, V., Robinson, T. D., Mahadevan, S., ... & Deshpande, R. 2013, *ApJ*, *765*, 2, 131.

Kopparapu, R. K., Ramirez, R. M., SchottelKotte, J., Kasting, J. F., Domagal-Goldman, S., & Eymet, V. 2014, *ApJL*, *787*, 2, L29.

Leconte, J., Forget, F., Charnay, B., Wordsworth, R., & Pottier, A. (2013). *Nature*, *504*(7479), 268-271.

Mischna, M. A., Kasting, J. F., Pavlov, A., & Freedman, R. 2000, *Icarus*, *145*, 2, 546-554.

Montési, L. G., & Zuber, M. T. (2003). *JGR:Planets*, *108*(E6).

Pierrehumbert, R., & Gaidos, E. 2011, *ApJL*, *734*, 1, L13.

Ramirez, R. M., Kopparapu, R., Zugger, M. E., Robinson, T. D., Freedman, R., & Kasting, J. F. 2014, *NatGe*, *7*, 1, 59-63.

Ramirez, R. M., & Kaltenegger, L. 2014, *ApJL*, *797*, 2, L25.

Ramirez, R. M., & Kaltenegger, L. 2016, *ApJ*, *823*, 1, 6.

Ramirez, R.M. 2016, *NatGe*, *9*, 6, 413-414.

Scaillet, B., & Gaillard, F. 2011, *Nature*, *480*, 7375, 48-49.

Seager, S., Bains, W., & Hu, R. 2013, *ApJ*, *777*, 2, 95.

Stanley, B. D., Hirschmann, M. M., & Withers, A. C. 2011, *Geochimica et Cosmochimica Acta*, *75*, 20, 5987-6003.

Tian, F., Toon, O. B., Pavlov, A. A., & De Sterck, H. 2005, *Science*, *308*, 5724, 1014-1017.

Trail, D., Watson, E. B., & Tailby, N. D. 2011, *Natur*, *480*, 7375, 79-82.

Stevenson, D. J. 1999, *Natur*, *400*, 6739, 32-32.

Wade, J., & Wood, B. J. 2005, *E&PSL*, *236*, 1, 78-95.

Walker, J. C. 1977, *New York: Macmillan, and London: Collier Macmillan, 1977, 318 pp.*

Wood, B. J., Walter, M. J., & Wade, J. 2006, *Natur*, *441*, 7095, 825-833.

Wordsworth, R. 2012, *Icarus*, *219*, 1, 267-273.

Wordsworth, R., & Pierrehumbert, R. 2013, S*cience*, *339*, 6115, 64-67.

Wordsworth, R., Kalugina, Y., Lokshtanov, S., et al. (2017). *GeoRL*. doi: 10.1002/2016GL071766

Zsom, A., Kaltenegger, L., & Goldblatt, C. (2012). *Icarus*, *221*(2), 603-616.